# Near-field infrared nanospectroscopy reveals guest confinement in metal-organic framework single crystals

Annika F. Möslein[a], Mario Gutiérrez[a], Boiko Cohen[b] and Jin-Chong Tan[a]*

[a]Multifunctional Materials & Composites (MMC) Laboratory, Department of Engineering Science, University of Oxford, Parks Road, Oxford OX1 3PJ, United Kingdom.

[b]Departamento de Química Física, Facultad de Ciencias Ambientales y Bioquímica, and INAMOL, Universidad de Castilla-La Mancha, Avenida Carlos III, S.N., 45071 Toledo, Spain.

*jin-chong.tan@eng.ox.ac.uk

**Abstract**

Metal-organic frameworks (MOFs) can provide exceptional porosity for molecular guest encapsulation useful for emergent applications in sensing, gas storage, drug delivery and optoelectronics. Central to the realisation of such applications however is the successful incorporation of a functional guest confined within the host framework. Here we demonstrate, for the first time, the feasibility of scattering-type scanning near-field optical microscopy (s-SNOM) and nano-Fourier transform infrared (nanoFTIR) spectroscopy, in concert with density functional theory (DFT) calculations to reveal the vibrational characteristics of the Guest@MOF systems. Probing individual MOF crystals, we pinpoint the local molecular vibrations and thus, shed new light on the host-guest interactions at the nanoscale. Our strategy not only confirms the successful encapsulation of luminescent guest molecules in the porous host framework in single crystals, but further provides a new methodology for nanoscale-resolved physical and chemical identification of wide-ranging framework materials and designer porous systems for advanced applications.

**INTRODUCTION**

Metal-organic frameworks (MOFs), characterised by their crystalline hybrid structure, are constructed from metal clusters and organic linkers *via* self-assembly at the molecular level. MOFs exhibit remarkably large internal surface areas, far exceeding those found in conventional porous materials such as zeolites and carbon black.[1] Merging the hybrid nature of MOFs with the ability to precisely tailor the characteristics of the pore yields multifunctional properties, boosting their deployment in emerging technologies ranging from gas storage and catalysis to luminescence, dielectrics, drug delivery and sensors.[2-6] Due to their potential in prospective optoelectronic and sensing technologies, the research interest in luminescent MOFs has intensified towards accomplishing MOF-based devices for real-world applications.[7-9] In this context, the encapsulation of "guest" functional molecules into the "host" MOF pores is a versatile strategy to engineer the Guest@MOF composite materials with tuneable physicochemical properties arising from host-guest interactions.[10, 11]

There are, however, outstanding challenges to be addressed to achieve the full potential of Guest@MOF systems. Particularly, it is very plausible that during the *in situ* synthesis or *ex situ* infiltration process, the guest molecules are adsorbed onto the external surfaces of MOFs instead of truly being encapsulated inside the pores. Unambiguously proving the latter is not a trivial task. Herein, we confirm – on single crystals – the fundamental encapsulation of luminescent guest molecules (fluorophores) into the framework structures of the MOF host material. Our nanoscale multimodal approach combines fluorescence lifetime imaging microscopy (FLIM) with precise determination of the vibrational dynamics of individual crystals, employing the scattering-type scanning near-field optical microscopy (s-SNOM) integrated with nano-Fourier transform infrared (nanoFTIR) spectroscopy, circumventing the diffraction limit of light.[12] The combination of the latter local-scale techniques enables us to perform single-crystal imaging and nanoscale chemical characterisation by simultaneously measuring topography and infrared-active vibrational modes, which yields spectral information spatially resolved down to 20 nm.[13] While interference microscopy and scanning transmission electron microscopy have been used to study crystal diversity with time-resolution[14] and heterodomains of MOFs,[15] respectively, s-SNOM measurements could surpass the spatial resolution of these techniques. The s-SNOM setup employed in this work is based on an atomic force microscope (AFM), where a platinum-coated cantilever tip serves as a topographical and near-field optical probe simultaneously (Fig.1a). Upon illumination, the probe induces an evanescent near-field by acting as a nanoscale light confiner, enhancer and scatterer - key to obtain wavelength-independent resolution (Fig.1b).[16] As the tip polarises the sample, the optical near-field interaction between the metallic tip and the sample modifies the elastically scattered light. Interferometric detection provides sensitivity to measure the sample's permittivity at a resolution comparable to the dimension of the tip apex.[16] When this signal is Fourier transformed, sample-specific FTIR spectra with a spatial resolution of down to 20 nm are obtained (hereafter we refer to these length scales as "local").[13, 17]

For the nanoFTIR method, the tip is operating in tapping mode at its mechanical resonance frequency ($\Omega$ = 250 kHz) under illumination from a tuneable broadband IR laser. Demodulating the detector signal at higher harmonics of the tip oscillation frequency ($n\Omega$) extracts the near-field interaction from background contributions.[18] Once normalised to a reference signal, this gives the complex-valued near-field contrast, whose imaginary part defines the local nanoFTIR absorption (further details in SI sections 1-2). s-SNOM imaging is achieved *via* a monochromatic irradiation source instead of a broadband laser. In this case, the illumination wavelength is tuned close to an absorption band of interest to map the material surface on a 2D areal scan. Analogous to nanoFTIR, the scattered light is detected and deconvoluted at higher harmonics of the tip frequency to record the local optical properties through near-field contributions. Optical amplitude and phase now indicate regions of the sample's reflection and absorption at the specific wavelength to derive contrast images with nanoscale resolution.

In this work, we first demonstrate the efficacy of the near-field optical techniques for the physical and chemical characterisation of MOF single crystals (*ca*. 100s nm – 1 $\mu$m). As a proof of concept, we measure nanoFTIR absorption spectra from individual crystals of zeolitic imidazolate framework ZIF-8 [$Zn(mIM)_2$; mIM = 2-methylimidazolate],[19] which represents a prototypical imidazole-based MOF with sodalite cage topology. We compare the near-field spectra with far-field FTIR measurements of a bulk sample, and with the theoretical spectra of ZIF-8. Subsequently, we demonstrate how to confirm the nanoscale confinement of luminescent guest molecules, such as rhodamine B (RhB) or fluorescein, encapsulated in ZIF-8 and UiO-66, the latter is a prototype of a highly stable zirconium-based framework [$Zr_6O_4(OH)_4(BDC)_6$; BDC = benzene-1,4-dicarboxylate].[20] Specifically, we study the RhB@ZIF-8, RhB@UiO-66, and fluorescein@UiO-66 composite systems, finally validating the Guest@MOF concept.

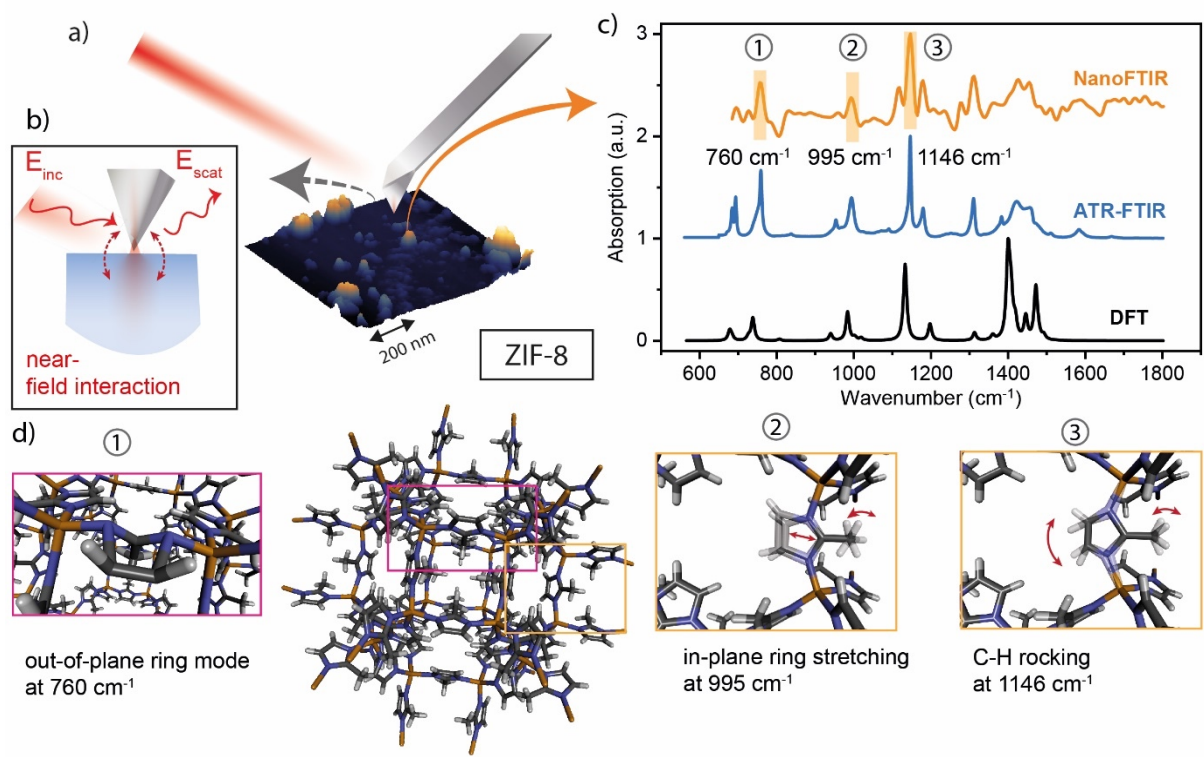

**Figure 1: Near-field optical spectroscopy of individual ZIF-8 nanocrystals.** (a) Representation of the setup of the s-SNOM measurement stage: the illuminated AFM tip generates a nanofocus on the sample. (b) The near-field interaction between the tip and the sample changes the scattered light from which the local optical properties of the sample are derived. (c) Mid-IR spectra of ZIF-8 crystals obtained *via* nanoFTIR and ATR-FTIR measurements, compared with the DFT calculations. The DFT spectrum was shifted by a factor of 0.97 to better match the experimental measurements.[21] (d) Characteristic vibrational modes of the ZIF-8 crystal structure: 1) out-of-plane deformation of the mIM ring at 760 cm$^{-1}$, 2) in-plane stretching of the mIM ring at 995 cm$^{-1}$ and 3) C-H rocking of the mIM linker at 1146 cm$^{-1}$.

**RESULTS AND DISCUSSIONS**

**IR spectroscopy of MOF single crystals**

We first demonstrate that the described nanoFTIR method can be employed to probe individual MOF-type crystals. Herein, we compare the vibrational spectra of near-field nanoFTIR experimental measurements of ZIF-8 crystals with far-field attenuated total reflection (ATR-FTIR) measurements, and these experiments against *ab initio* quantum mechanical calculations from density functional theory (DFT). The results are shown in Fig.1c. In the mid-IR region, the characteristic peaks at 760 cm$^{-1}$, 995 cm$^{-1}$ and 1146 cm$^{-1}$ are all fully resolved in the nanoscale measurements. Guided by DFT calculations of ZIF-8,[22] we assign these vibrational bands to the out-of-plane and the in-plane ring stretching modes of the mIM linker, as well as the rocking mode of its C-H bonds, respectively, as illustrated in Fig.1d.

Similarly, the in-plane ring bending modes of C-N bonds at 1116 cm$^{-1}$ and 1311 cm$^{-1}$, as well as the symmetric C-N stretching mode at 1445 cm$^{-1}$ are present in the nanoFTIR spectra.

Indeed, the matching IR spectra demonstrate the capability of nanoFTIR spectroscopy to characterise a MOF-type single crystal; nonetheless, this near-field technique yields minor changes in the IR spectrum. It is worth mentioning that our DFT calculation assumes a defect-free periodic crystal and neglects anharmonicity, while the far-field ATR-FTIR method measures the averaged response of a bulk (polycrystalline) powder material. In contrast, a local scale characterisation of individual single crystals is achieved by leveraging nanoFTIR. Here, with a probing depth of ~20 nm, surface effects might influence the IR spectrum. Close to the crystal boundary, where the framework symmetry is lost, some functional groups are surrounded by voids and air instead of their atomic neighbours assumed in a periodic crystalline structure. Such changes in the atomic environment will affect the strength of the bonds, altering their vibrational frequencies, and as a result, additional peaks and broader peak shapes are observed (Fig.1c). Although subtle alterations in the IR spectrum are observable, they are explained by the nature of experimental surface measurements. In fact, nanoFTIR spectroscopy yields good agreement with established FTIR and DFT methods, thus allowing direct chemical recognition of MOFs through standard FTIR databases, furthermore it has the unique advantage to directly measure a local IR spectrum reflecting the complex nature of a single MOF crystal whose size is of the order of ~100 nm.

**Host-guest interactions at the nanoscale**

Local probing enables a deeper examination of the host-guest interactions in Guest@MOF systems. Crucially, one might ask whether the guest molecule is actually incorporated in the pores of the framework or adsorbed only to its external surfaces, which, to date, has been a major challenge. Utilising nanoFTIR and s-SNOM imaging, for the first time, give us the unique opportunity to chemically pinpoint the interaction of the guest molecule at the nanoscale with the MOF host.

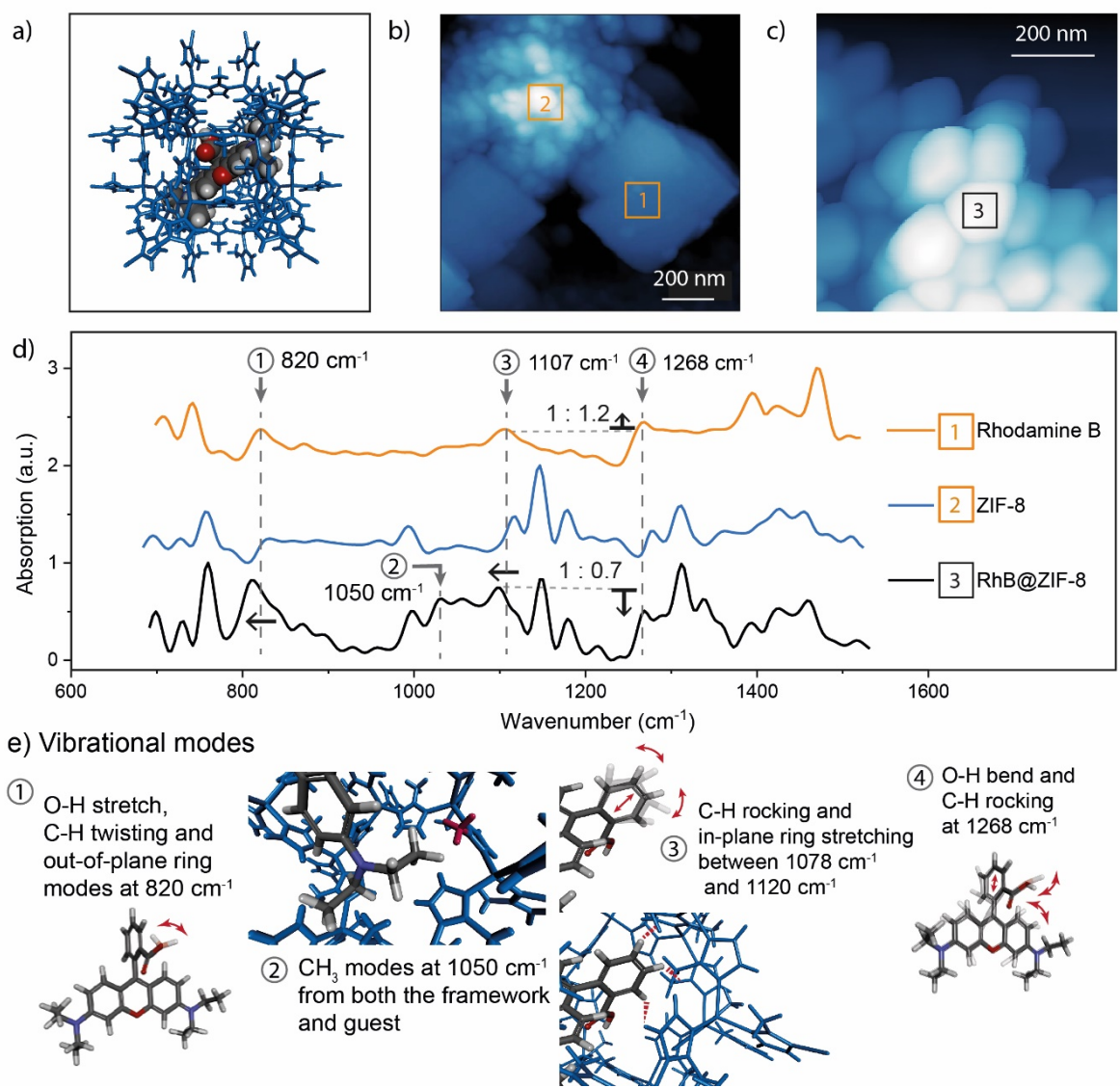

**Figure 2: Vibrational analysis of rhodamine B (RhB) and ZIF-8 *via* nanoFTIR and DFT calculations.** (a) Schematic representation of the RhB@ZIF-8 composite, depicting a RhB guest molecule being encapsulated in the pore of the ZIF-8 host framework (in blue). (b) AFM image of the as-synthesised sample containing two distinctive phases: (1) RhB and (2) ZIF-8 showing the positions where IR spectra were recorded. (c) AFM image of a single-phase sample of ZIF-8 nanocrystals adsorbing RhB. (d) nanoFTIR spectra determined at the designated locations on the AFM image. (e) Vibrational modes of the RhB@ZIF-8 composite illustrating the interactions between the ZIF-8 host framework and the RhB guest.

Here we consider an example Guest@MOF system, termed RhB@ZIF-8, comprising the luminescent rhodamine B (RhB) guest in ZIF-8 host, resulting in a fluorescent material potentially useful for optoelectronics and photonic sensors.[23, 24] However, the adsorption of RhB in pre-assembled ZIF-8 crystals only leads to surface interactions, since the size of the RhB molecule (13.5 Å) exceeds the window aperture of the ZIF-8 pore (3.4 Å), thus hindering guest infiltration.[25] Therefore, we applied an *in situ* encapsulation of RhB molecules during ZIF-8 formation to synthesise the RhB@ZIF-8 composite (see Methods and Fig.S3). Albeit, to isolate the RhB@ZIF-8 as synthesised, several washing steps were performed, AFM imaging revealed different phases with round nanocrystals and micron-sized blocks as their distinct morphologies (Fig.2b). Equally, local probing of the features with nanoFTIR spectroscopy yields two significantly different IR spectra. First, the cuboidal block is identified as pure RhB crystals through comparison with the DFT-simulated IR peaks (at 820 $cm^{-1}$, 1107 $cm^{-1}$, 1268 $cm^{-1}$, 1470 $cm^{-1}$ and 1576 $cm^{-1}$, calculated by Gaussian). Secondly, the frequencies measured at the individual round nanocrystals are matching the IR spectrum of pristine ZIF-8. Consequently, disparate regions of RhB and ZIF-8 are clearly distinguishable by nanoFTIR, where the topography and chemical fingerprint for each constituent material are identifiable without any indication of molecular interaction.

In the third region (Fig.2c), the block crystals of RhB are absent, and the topography resembles the ZIF-8 nanocrystals. Interestingly, probing the individual crystals in this region yields an IR spectrum with characteristic features of both constituent materials (Fig.2d). Although far-field FTIR measurements may equally lead to a combined IR spectrum, it is worth emphasising that this is due to the averaged probing over the bulk polycrystalline material without discriminating between the constituent materials present at the local scale. At the measured spot size of 20 nm, in contrast, the superposition of the spectra reveals the concurrent presence of RhB and ZIF-8, leading to the conclusion that the RhB interacts with the ZIF-8 framework due to two reasons. First, all characteristic modes associated with the host framework were identified in the IR spectrum. Secondly, the vibrational bands assigned to RhB have been modified, revealing host-guest interactions. Considering the O-H group, for instance, with its stretch mode at 820 $cm^{-1}$ experiencing a small shift to lower energies, indicates an interaction between the O-H group of the guest molecule and the host framework, possibly with the Zn-atoms at defect sites.[25, 26] Similarly, the bending mode of the O-H group at 1268 $cm^{-1}$ reveals a change in relative intensity with respect to the peak at 1107 $cm^{-1}$. We found that the $CH_3$ vibrations in the region ~1050 $cm^{-1}$, are reinforced due to superposition. The presence of both materials at a nanoscale spot demonstrates the interaction of RhB with the framework, however, this might be influenced by surface adsorption effects. We also observed reduced energy of the phenyl ring stretching modes in RhB at 1107 $cm^{-1}$ and 1118 $cm^{-1}$, which may indicate guest encapsulation. As the dimensions of the RhB molecule almost completely fill the pore of ZIF-8, upon confinement, the free-space vibrations of the trapped molecules are suppressed.

**Confirming the confinement of RhB in the ZIF-8 framework**

One reason for the observed simultaneous presence of guest and framework material at a spot size of 20 nm may be attributed to surface adhesion. To eliminate any residual RhB from adhering on the surface of ZIF-8 nanocrystals, the synthesised RhB@ZIF-8 material was subjected to a second, more thorough washing process (see Methods). If, despite any dye material being removed from the surface, the luminescent properties of RhB are still observable, then the luminescent guests shall be incorporated. On this basis, the material was thoroughly washed until the sample resembled the white colour of pristine ZIF-8 rather than the characteristic pink of RhB (Fig.3c). As verification, the supernatant of the final washing step was characterised by UV absorption and fluorescence spectroscopy; and since no signal of the dye was obtained, the thorough washing is confirmed (Fig.S12). Both the as-synthesised RhB@ZIF-8 (Fig.3d) and the material obtained after thorough washing (Fig.3e) were initially probed using s-SNOM imaging at the illumination wavelength of 1146 $cm^{-1}$, a characteristic peak of ZIF-8 (Fig.3a), to confirm the presence of ZIF-8 nanocrystals. Upon illumination, only the ZIF-8 crystals absorb IR radiation, represented by the red colour in the optical contrast image (Fig.3b), whilst the RhB block appears transparent like the gold substrate (blue colour). Conversely, excitation at 1470 $cm^{-1}$ leads to the inverted contrast due to the strong absorbance of the excited C=C modes of RhB (Fig.3b). Since the surface of the nanocrystals isolated prior to exhaustive washing exhibits regions where the distinctive wavelength for RhB is absorbed, we can unambiguously conclude that RhB is still attached to the surface. After thorough washing, however, no trace of RhB can be found on the surface of the ZIF-8 nanocrystals. The same phenomenon was observed at larger scan areas, where, for clarity, the optical phase contrast is normalised to demonstrate the excess RhB on the RhB@ZIF-8 material before washing. Note that the red contrast in the thoroughly washed material can be assigned to a shadowing effect at the crystal edges (not originating from intrinsic optical properties of the material), because only the areas surrounding the nanocrystals absorb light.

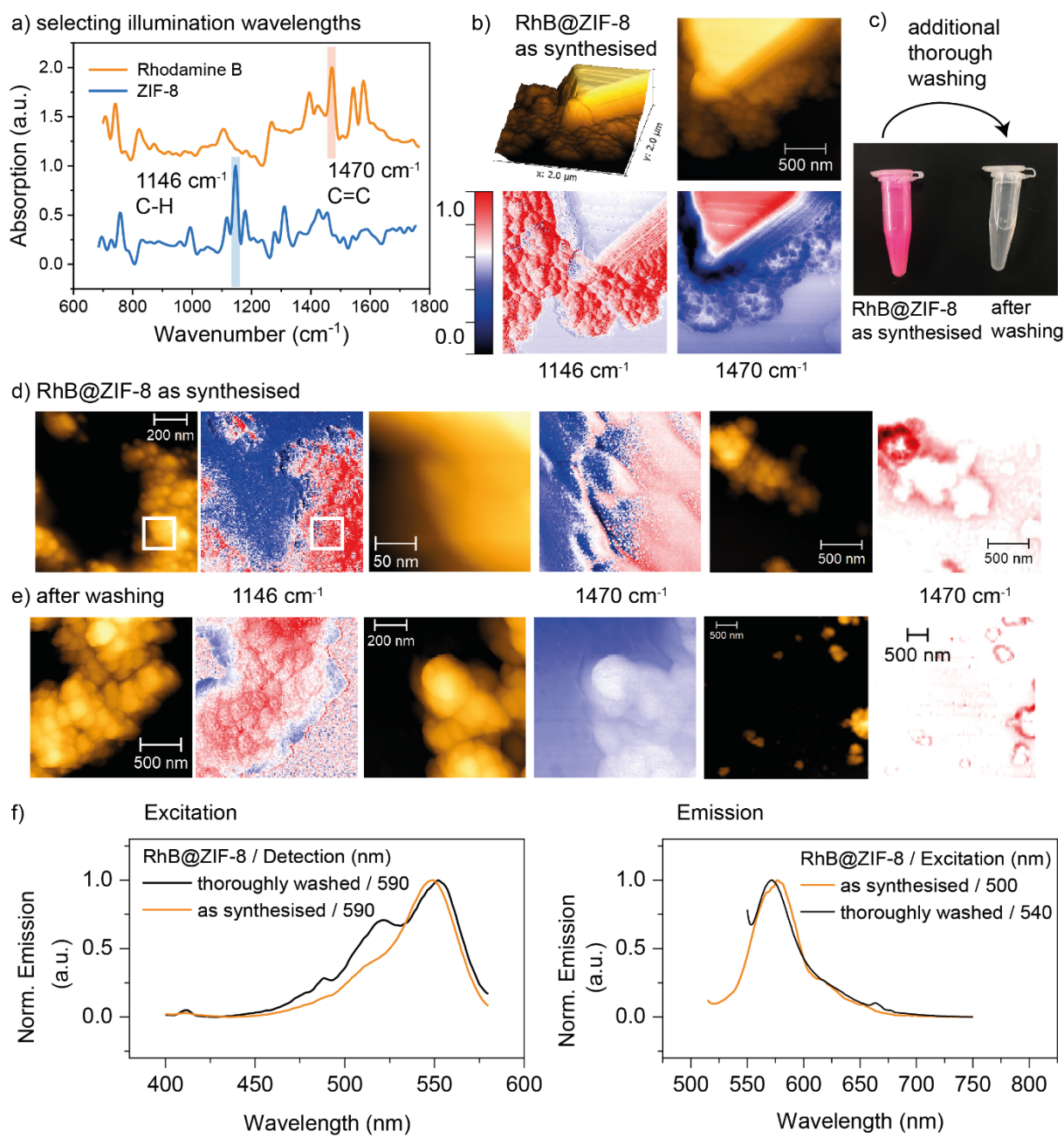

**Figure 3: s-SNOM imaging of RhB@ZIF-8.** (a) The illumination source was tuned to the characteristic peaks of RhB and ZIF-8. (b) Validation of the contrast images with known, distinguishable material distribution. (c) Samples before and after thorough washing, viewed under daylight. (d) AFM and near-field optical phase imaging of as-synthesised RhB@ZIF-8 with illumination at 1146 cm$^{-1}$ confirming the presence of ZIF-8, and illumination at 1470 cm$^{-1}$ still revealing RhB on the surface. (e) Near-field optical phase imaging of RhB@ZIF-8 after thorough washing, verifying the complete removal of residual RhB from the sample surface. (f) Normalised excitation and emission spectra of the as-synthesised and thoroughly washed RhB@ZIF-8 (dispersed in acetone) confirming the presence of RhB in both samples. Excitation and detection wavelengths are indicated in the figure.

s-SNOM imaging thus indicates the absence of RhB adhered onto the surface of the ZIF-8 nanocrystals (after thorough washing), with the removal of any excess guest material from the sample further confirmed by SEM imaging (Fig.S8). Remarkably, the excitation and emission spectra of the thoroughly washed RhB@ZIF-8 composite show the characteristic bands of RhB (Fig.3f). Fluorescence spectroscopic data reveal the presence of the luminescent dye within ZIF-8, which leads to the conclusion that the guest molecule is successfully encapsulated within the framework. This is a strong evidence verifying the formation of the RhB@ZIF-8 composite.

**Revealing the guest encapsulation in UiO-66**

To demonstrate the efficacy of the developed methodologies for determining the confinement of guest molecules in MOFs, we have employed two novel Guest@MOF systems as case studies: RhB@UiO-66 and fluorescein@UiO-66 (Fig.4 and Fig.S4). We found that the as-synthesised RhB@UiO-66 still exhibits traces of excess guest material on the surface of the nanocrystals probed *via* nanoFTIR (Fig.4a). Here, the peaks correlating to the characteristic modes of RhB (in the region between 820 $cm^{-1}$ and 1268 $cm^{-1}$) are visible in the nanoFTIR spectrum in addition to the peaks of the pristine UiO-66 crystals (Fig.4a). An exhaustive washing process – again, confirmed by probing the supernatant with UV absorption and fluorescence spectroscopy (Fig.S12) – eliminates any unconfined RhB and thus, the IR spectrum probed at nanoscale resolution on the surface of the crystals only reveals the representative peaks of UiO-66 (Fig.4a). Turning to the fluorescein@UiO-66 system, likewise, after thoroughly removing any excess guest material, the nanocrystals were identified *via* nanoFTIR as UiO-66 without any sign of fluorescein on the surface (Fig.4b).

AFM and SEM imaging of the thoroughly washed samples demonstrate the homogenous crystals without any excess guest material on the sample (Fig.4c-d, Fig.S9). To further confirm the absence of the guest adhering on the surface, not only local point spectra but rather the surface of the fluorescein@UiO-66 crystals were examined using s-SNOM imaging (Fig.5b). Hereby, tuning the monochromatic irradiation source to the pronounced peak of UiO-66 at 1408 $cm^{-1}$ illustrates the strong absorbance and reflectance of the crystals in comparison with the substrate. Little contrast, on the other hand, was obtained from illumination at 1325 $cm^{-1}$, a characteristic peak of fluorescein. Although some absorbance can be observed on the edges of the crystals, again, this can be attributed to noise and shadowing because the same phenomenon emerges at a reference wavelength (1440 $cm^{-1}$), where neither fluorescein nor the UiO-66 material absorbs. Hence, the absence of any guest material adsorbed on the crystal surface is confirmed. As opposed to the surface techniques, where no signal of the guest molecules is detected, the characteristic emission and excitation bands of RhB and fluorescein are clearly identified for RhB@UiO-66 and fluorescein@UiO-66 composites when measuring their photophysical properties *via* fluorescence spectroscopy (Fig.5d-e).

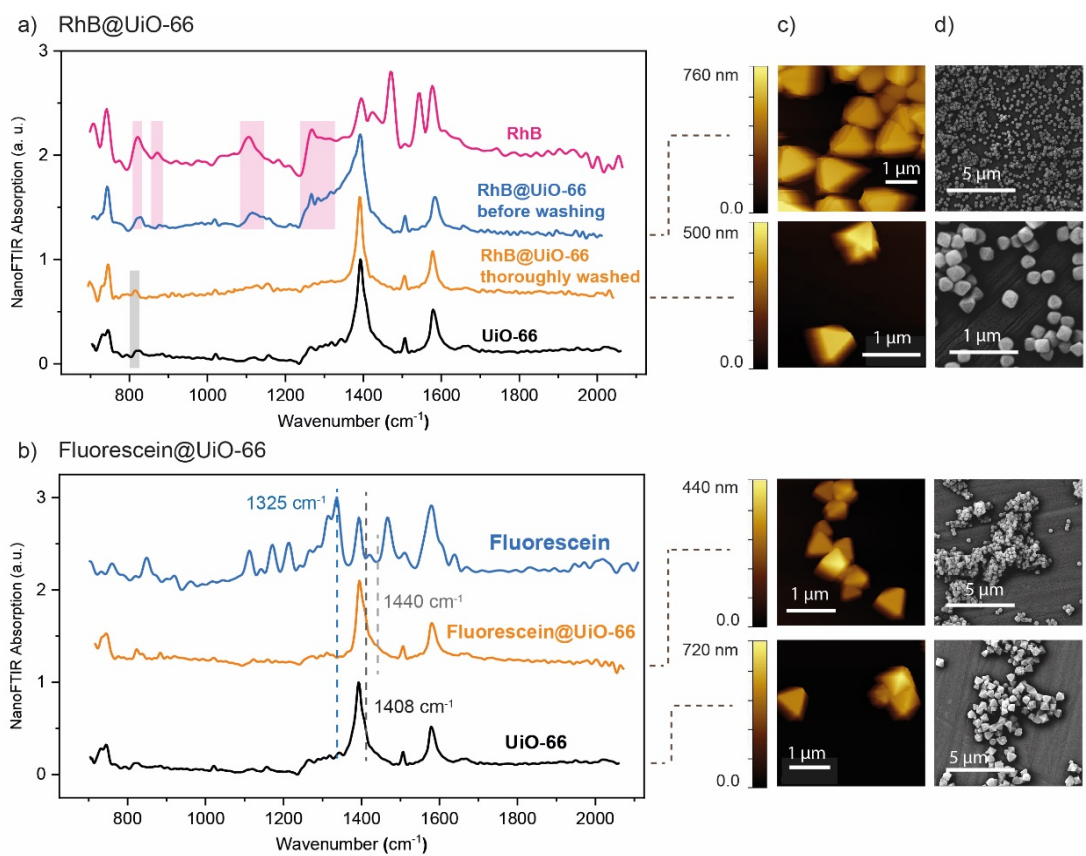

**Figure 4: Near-field nanospectroscopy of UiO-66 single crystals with the encapsulated guest RhB and fluorescein, respectively.** (a) Near-field IR absorption spectra of RhB, UiO-66, and as-synthesised RhB@UiO-66 composite (without washing the sample) and after thorough washing. (b) The nanoFTIR spectrum of fluorescein@UiO-66 measured on thoroughly washed nanocrystals revealing the absence of fluorescein on the sample surface. Distinguishable wavelengths are selected for subsequent s-SNOM imaging. (c) AFM and (d) SEM images of UiO-66, RhB@UiO-66 and fluorescein@UiO-66 crystals.

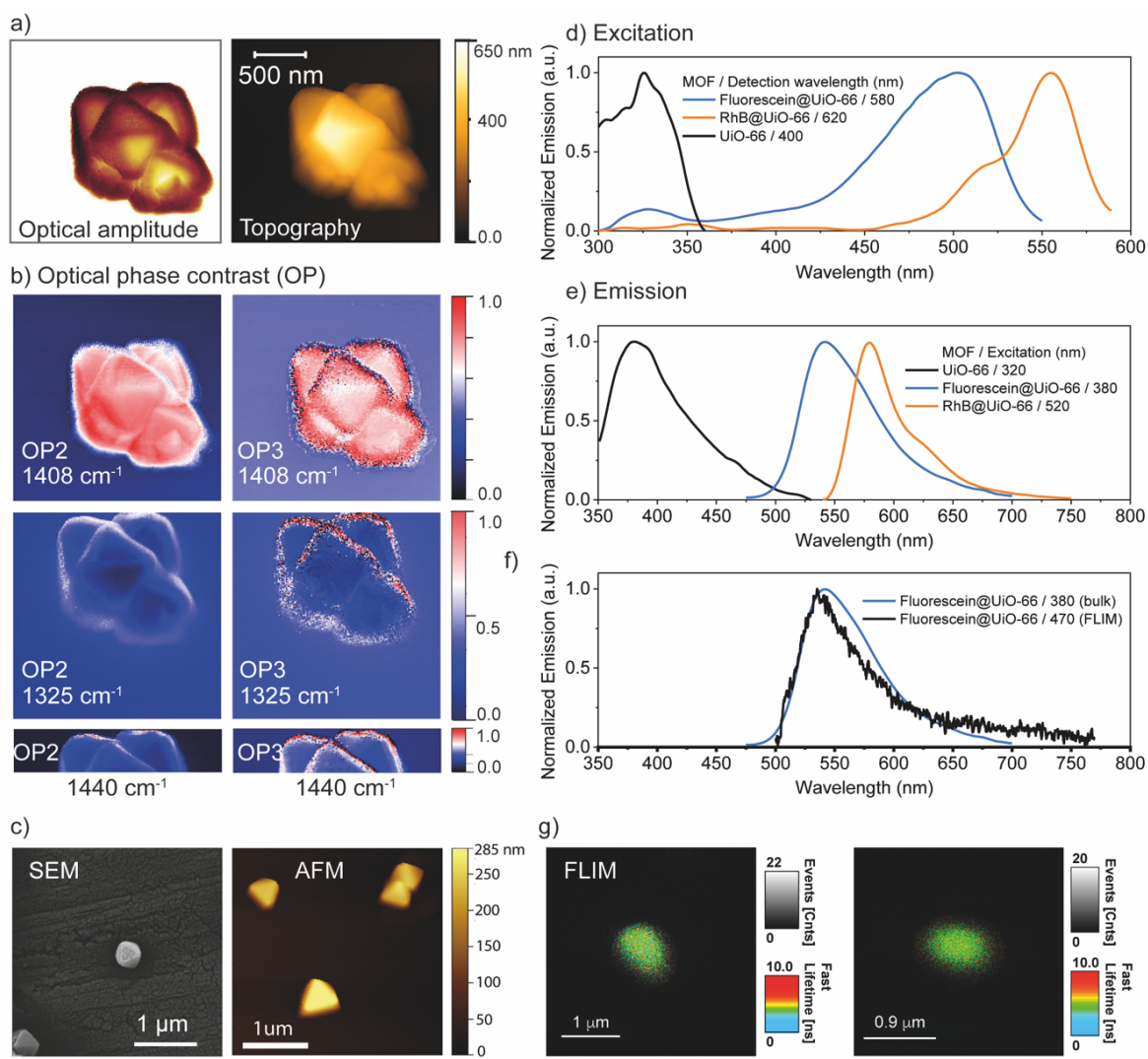

**Figure 5: Nanoscale imaging confirming guest encapsulation in UiO-66.** (a) Topography and optical amplitude imaging demonstrating the typical size and morphology of the crystals. (b) s-SNOM imaging proves that there are no traces of fluorescein on the surface of the UiO-66 crystals. The optical phase images were obtained *via* illumination at the major absorption peaks of UiO-66 (1408 cm$^{-1}$) and fluorescein (1325 cm$^{-1}$), and at a reference position (1440 cm$^{-1}$), as shown in figure 4b. (c) SEM and AFM imaging of the individual UiO-66 crystals. (d-e) Excitation and emission spectra of UiO-66 and thoroughly washed Guest@UiO-66 powder samples. Excitation and detection wavelengths are indicated in the figures. (f) Emission spectrum of a single crystal of fluorescein@UiO-66 obtained with a confocal fluorescence microscope compared with the spectrum of powder sample revealing the presence of fluorescein. (g) Fluorescence lifetime imaging microscopy (FLIM) confirming the encapsulation of fluorescein in UiO-66 single crystals, measured under an excitation wavelength of 470 nm.

Finally, to further corroborate the encapsulation of the fluorophores at a single crystal level, the fluorescence lifetime image (FLIM) of fluorescein@UiO-66 was recorded using a confocal fluorescence microscope (see Methods). Fig.5g shows two single-crystal FLIMs, where clearly, the fluorescein molecules are homogeneously distributed within the UiO-66 crystal. Moreover, the emission spectrum (Fig.5f) of these crystals correlates well with the emission of bulk fluorescein@UiO-66, proving that the emission signal is generated by the fluorescein guest molecules. The mean lifetime of the single crystals is a monoexponential decay of $\tau$ = 3.77 ± 0.15 ns (Fig.S10) consistent with the lifetime of fluorescein in different solvents.[27-29] This result suggests that fluorescein molecules are isolated in the form of monomers when partitioned by the pores, as the formation of aggregates will give multiexponential lifetimes.[30] The homogenous distribution of fluorescein molecules in the form of isolated monomers, alongside with the non-observation of fluorescein on the surface of UiO-66 crystals derived from the nanoFTIR experiments, unequivocally proves the encapsulation of the fluorophores into the MOF pores.

**CONCLUSIONS**

In summary, this first study of individual ZIF-8 and UiO-66 nanocrystals using near-field optical nanospectroscopy yields new understanding about MOFs and their host-guest interactions. We show the capability of the nanoFTIR methodology to characterise individual MOF-type crystals by comparison with established far-field techniques and *ab initio* theoretical calculations. Taking a step further towards precisely unravelling the physical and chemical properties of MOFs, we present the first near-field spectroscopic evidence of host-guest interactions by locally characterising both constituent materials at a 20 nm irradiation spot. A detailed analysis of the nanocrystals leveraging s-SNOM imaging confirms the absence of any guest material adsorbed on the surface (after thorough washing), while the photophysical properties of the luminescent molecules are still observable. We conclude that the guest molecule is confined in the framework and thus, evidencing the Guest@MOF concept. Our findings provide the groundwork to gain a fundamental understanding of the host-guest interactions underpinning hybrid framework materials, further inviting the examination of other porous materials: covalent organic frameworks, 2D nanosheets,[11] mixed-matrix membranes,[31] and polycrystalline films for sensing devices.[32, 33] Confirming the guest encapsulation paves the way to controlling properties at the nanoscale — the key to engineer a multifunctional platform for nanotechnological applications.

**SUPPORTING INFORMATION**

Synthetic procedures in detail, methods used for materials characterisation (PXRD, NanoFTIR spectroscopy, s-SNOM imaging, fluorescence spectroscopy, FLIM, DFT calculations), further details on measuring the local infrared absorption, additional figures of material characterisation (XRD, AFM, SEM, FLIM).

**ACKNOWLEDGEMENTS**

A.F.M. thanks the Oxford Ashton Memorial scholarship for a DPhil studentship award. J.C.T. and A.F.M. thank the EPSRC Grant No. EP/N014960/1 and ERC Consolidator Grant under the grant agreement 771575 (PROMOFS) for funding. We would like to acknowledge the use of the University of Oxford Advanced Research Computing (ARC) facility in carrying out this work (http://dx.doi.org/10.5281/zenodo.22558). We are grateful to the Research Complex at Harwell (RCaH) for the provision of advanced materials characterisation facilities. A.F.M. would like to thank Dr Cyril Besnard and Prof. Alexander Korsunsky for their help with SEM imaging, and neaspec GmbH for the opportunity to perform s-SNOM imaging in Haar, Germany. We also thank Dr Abhijeet Chaudhari for the provision of the RhB@ZIF-8 samples.

References

1. Furukawa, H.; Cordova, K. E.; O'Keeffe, M.; Yaghi, O. M. The chemistry and applications of metal-organic frameworks. *Science* **2013,** 341 (6149), 1230444.
2. Zhu, L.; Liu, X. Q.; Jiang, H. L.; Sun, L. B. Metal-Organic Frameworks for Heterogeneous Basic Catalysis. *Chem. Rev.* **2017,** 117 (12), 8129-8176.
3. Horcajada, P.; Gref, R.; Baati, T.; Allan, P. K.; Maurin, G.; Couvreur, P.; Ferey, G.; Morris, R. E.; Serre, C. Metal-organic frameworks in biomedicine. *Chem. Rev.* **2012,** 112 (2), 1232-1268.
4. Li, B.; Wen, H. M.; Zhou, W.; Chen, B. Porous Metal-Organic Frameworks for Gas Storage and Separation: What, How, and Why? *J. Phys. Chem. Lett.* **2014,** 5 (20), 3468-3479.
5. Lustig, W. P.; Mukherjee, S.; Rudd, N. D.; Desai, A. V.; Li, J.; Ghosh, S. K. Metal-organic frameworks: functional luminescent and photonic materials for sensing applications. *Chem. Soc. Rev.* **2017,** 46 (11), 3242-3285.
6. Ryder, M. R.; Zeng, Z.; Titov, K.; Sun, Y.; Mahdi, E. M.; Flyagina, I.; Bennett, T. D.; Civalleri, B.; Kelley, C. S.; Frogley, M. D.; Cinque, G.; Tan, J.-C. Dielectric Properties of Zeolitic Imidazolate Frameworks in the Broad-Band Infrared Regime. *J. Phys. Chem. Lett.* **2018,** 9 (10), 2678-2684.
7. Kaur, H.; Sundriyal, S.; Pachauri, V.; Ingebrandt, S.; Kim, K.-H.; Sharma, A. L.; Deep, A. Luminescent metal-organic frameworks and their composites: Potential future materials for organic light emitting displays. *Coord. Chem. Rev.* **2019,** 401, 213077.
8. Stassen, I.; Burtch, N.; Talin, A.; Falcaro, P.; Allendorf, M.; Ameloot, R. An updated roadmap for the integration of metal-organic frameworks with electronic devices and chemical sensors. *Chem. Soc. Rev.* **2017,** 46 (11), 3185-3241.
9. Gutiérrez, M.; Martín, C.; Van der Auweraer, M.; Hofkens, J.; Tan, J. C. Electroluminescent Guest@MOF Nanoparticles for Thin Film Optoelectronics and Solid‐State Lighting. *Adv. Opt. Mater.* **2020**, 2000670.
10. Allendorf, M. D.; Foster, M. E.; Leonard, F.; Stavila, V.; Feng, P. L.; Doty, F. P.; Leong, K.; Ma, E. Y.; Johnston, S. R.; Talin, A. A. Guest-Induced Emergent Properties in Metal-Organic Frameworks. *J. Phys. Chem. Lett.* **2015,** 6 (7), 1182-1195.
11. Chaudhari, A. K.; Kim, H. J.; Han, I.; Tan, J. C. Optochemically Responsive 2D Nanosheets of a 3D Metal-Organic Framework Material. *Adv. Mater.* **2017,** 29 (27), 1701463.
12. Knoll, B.; Keilmann, F. Near-field probing of vibrational absorption for chemical microscopy. *Nature* **1999,** 399, 134-137.
13. Huth, F.; Govyadinov, A.; Amarie, S.; Nuansing, W.; Keilmann, F.; Hillenbrand, R. Nano-FTIR absorption spectroscopy of molecular fingerprints at 20 nm spatial resolution. *Nano Lett.* **2012,** 12 (8), 3973-3978.
14. Saint Remi, J. C.; Lauerer, A.; Chmelik, C.; Vandendael, I.; Terryn, H.; Baron, G. V.; Denayer, J. F.; Karger, J. The role of crystal diversity in understanding mass transfer in nanoporous materials. *Nat. Mater.* **2016,** 15 (4), 401-6.
15. Collins, S. M.; Kepaptsoglou, D. M.; Butler, K. T.; Longley, L.; Bennett, T. D.; Ramasse, Q. M.; Midgley, P. A. Subwavelength Spatially Resolved Coordination Chemistry of Metal-Organic Framework Glass Blends. *J. Am. Ceram. Soc.* **2018,** 140 (51), 17862-17866.
16. Keilmann, F.; Hillenbrand, R. Near-field microscopy by elastic light scattering from a tip. *Phil. Trans. R. Soc. A* **2004,** 362 (1817), 787-805.
17. Zenhausern, F.; Martin, Y.; Wickramasinghe, H. K. Scanning interferometric apertureless microscopy: Optical imaging at 10 angstrom resolution. *Science* **1995,** 269, 1083-1085.
18. Govyadinov, A. A.; Amenabar, I.; Huth, F.; Carney, P. S.; Hillenbrand, R. Quantitative Measurement of Local Infrared Absorption and Dielectric Function with Tip-Enhanced Near-Field Microscopy. *J. Phys. Chem. Lett.* **2013,** 4 (9), 1526-1531.
19. Tan, J.-C.; Civalleri, B.; Lin, C.-C.; Valenzano, L.; Galvelis, R.; Chen, P.-F.; Bennett, T. D.; Mellot-Draznieks, C.; Zicovich-Wilson, C. M.; Cheetham, A. K. Exceptionally Low Shear Modulus in a Prototypical Imidazole-Based Metal-Organic Framework. *Phys. Rev. Lett.* **2012,** 108 (9), 095502.

20.     Cavka, J. H.; Jakobsen, S.; Olsbye, U.; Guillou, N.; Lamberti, C.; Bordiga, S.; Lillerud, K. P. A New Zirconium Inorganic Building Brick Forming Metal Organic Frameworks with Exceptional Stability. *J. Am. Chem. Soc.* **2008,** 130 (42), 13850–13851.
21.     Rauhut, G.; Pulay, P. Transferable Scaling Factors for Density Functional Derived Vibrational Force Fields. *J. Phys. Chem.* **1995,** 99 (10), 3093-3100.
22.     Ryder, M. R.; Civalleri, B.; Bennett, T. D.; Henke, S.; Rudic, S.; Cinque, G.; Fernandez-Alonso, F.; Tan, J. C. Identifying the role of terahertz vibrations in metal-organic frameworks: from gate-opening phenomenon to shear-driven structural destabilization. *Phys. Rev. Lett.* **2014,** 113 (21), 215502.
23.     Zhang, Y. Q.; Wu, X. H.; Mao, S.; Tao, W. Q.; Li, Z. Highly luminescent sensing for nitrofurans and tetracyclines in water based on zeolitic imidazolate framework-8 incorporated with dyes. *Talanta* **2019,** 204, 344-352.
24.     Chaudhari, A. K.; Tan, J. C. Dual‐Guest Functionalized Zeolitic Imidazolate Framework‐8 for 3D Printing White Light-Emitting Composites. *Adv. Opt. Mater.* **2020,** 8 (8), 1901912.
25.     Chin, M.; Cisneros, C.; Araiza, S. M.; Vargas, K. M.; Ishihara, K. M.; Tian, F. Rhodamine B degradation by nanosized zeolitic imidazolate framework-8 (ZIF-8). *RSC Adv.* **2018,** 8 (47), 26987-26997.
26.     Zhang, C.; Han, C.; Sholl, D. S.; Schmidt, J. R. Computational Characterization of Defects in Metal-Organic Frameworks: Spontaneous and Water-Induced Point Defects in ZIF-8. *J. Phys. Chem. Lett.* **2016,** 7 (3), 459.
27.     Zhang, X. F.; Zhang, J.; Liu, L. Fluorescence properties of twenty fluorescein derivatives: lifetime, quantum yield, absorption and emission spectra. *J. Fluoresc.* **2014,** 24 (3), 819-26.
28.     Magde, D.; Rojas, G. E.; Seybold, P. G. Solvent Dependence of the Fluorescence Lifetimes of Xanthene Dyes. *Photochem. Photobiol.* **1999,** 70 (5), 737-744.
29.     Martin, M. M.; Lindqvist, L. The pH dependence of fluorescein fluorescence. *J. Lumin.* **1975,** 10 (6), 381-390.
30.     Hungerford, G.; Benesch, J.; Mano, J. F.; Reis, R. L. Effect of the labelling ratio on the photophysics of fluorescein isothiocyanate (FITC) conjugated to bovine serum albumin. *Photochem. Photobiol. Sci.* **2007,** 6 (2), 152-158.
31.     Dechnik, J.; Sumby, C. J.; Janiak, C. Enhancing Mixed-Matrix Membrane Performance with Metal–Organic Framework Additives. *Cryst. Growth Des.* **2017,** 17 (8), 4467-4488.
32.     Yuan, H.; Li, N.; Linghu, J.; Dong, J.; Wang, Y.; Karmakar, A.; Yuan, J.; Li, M.; Buenconsejo, P. J. S.; Liu, G.; Cai, H.; Pennycook, S. J.; Singh, N.; Zhao, D. Chip-Level Integration of Covalent Organic Frameworks for Trace Benzene Sensing. *ACS Sens.* **2020,** 5 (5), 1474-1481.
33.     Stassen, I.; Styles, M.; Van Assche, T.; Campagnol, N.; Fransaer, J.; Denayer, J.; Tan, J.-C.; Falcaro, P.; De Vos, D.; Ameloot, R. Electrochemical Film Deposition of the Zirconium Metal–Organic Framework UiO-66 and Application in a Miniaturized Sorbent Trap. *Chem. Mater.* **2015,** 27 (5), 1801-1807.

*Supporting Information*

*for*

# Near-field infrared nanospectroscopy reveals guest confinement in metal-organic framework single crystals

Annika F. Möslein[a], Mario Gutiérrez[a], Boiko Cohen[b] and Jin-Chong Tan[a]*

[a]Multifunctional Materials & Composites (MMC) Laboratory, Department of Engineering Science, University of Oxford, Parks Road, Oxford OX1 3PJ, United Kingdom.

[b]Departamento de Química Física, Facultad de Ciencias Ambientales y Bioquímica, and INAMOL, Universidad de Castilla-La Mancha, Avenida Carlos III, S.N., 45071 Toledo, Spain.

*jin-chong.tan@eng.ox.ac.uk

# Contents

# 1 METHODS

## 1.1 SYNTHESIS PROTOCOLS

### 1.1.1 Synthesis of ZIF-8

Nanocrystals of ZIF-8 [$Zn(mIM)_2$; mIM = 2-methylimidazolate] were synthesised by dissolving 4.5 mmol of zinc nitrate hexahydrate ($Zn(NO_3)_2 \cdot 6H_2O$, 98%, Sigma-Aldrich) and 13.5 mmol 2-methylimidazole (98%, Sigma-Aldrich) in 60 mL of methanol, respectively. After combining the two clear solutions, the white colloidal solution was rigorously stirred for 1 min and then left to form the nanocrystals. After 60 min, nanocrystals were isolated by centrifugation at 8000 rpm for 10 min and, subsequently, the clear solution was substituted with fresh methanol. To remove any excess or unreacted metal salts and/or organic linkers, this washing process was repeated three times. Subsequently, the samples were drop casted onto gold substrates and dried at 80 °C for 4 hours in vacuum to remove any residual methanol as preparation for AFM characterisation.

### 1.1.2 Synthesis of RhB@ZIF-8

3 mmol of $Zn(NO_3)_2 \cdot 6H_2O$ was diluted in 20 mL of dimethylformamide (DMF). Similarly, 7.5 mmol of 2-methylimidazole was dissolved in 20 mL of DMF with addition of 7.5 mmol of triethylamine. The linker and the dye solution, the latter formed by 0.05 mmol (24 mg) of rhodamine B in 10 mL DMF, were mixed and later added to the metal ion solution. The material obtained was washed with 50 mL of fresh DMF, followed by two washing steps with acetone (termed RhB@ZIF-8 as synthesised). For additional, thorough washing, the material was diluted in 30 mL of fresh methanol under constant stirring for 60 min before centrifugation at 8000 rpm for 10 min to isolate the crystals. Subsequently, the same procedure was followed with fresh acetone. Both steps were alternated 5 times. Finally, the isolated nanocrystals were dried at 80 °C for 4 hours in vacuum.

### 1.1.3 Synthesis of UiO-66

Single crystals of UiO-66 [$Zr_6O_4(OH)_4(BDC)_6$; BDC = benzene-1,4-dicarboxylate] were prepared following a previously reported method.[1] Briefly, 0.4 mmol (93 mg) of zirconium chloride anhydrous ($ZrCl_4$, Fisher Scientific) and 0.4 mmol (67 mg, Fisher Scientific) of terephthalic acid were independently dissolved in 10 mL of DMF each. Then, 50 mmol (2.8 mL, Fisher Scientific) of glacial acetic acid were poured into the terephthalic acid solution and this mixture was transferred to a Pyrex bottle containing the $ZrCl_4$ solution. The final mixture was heated in an oven at 120 °C for 24 h. The white solid powder was thoroughly washed with DMF and MeOH, collected by centrifugation (8000 rpm), and dried at 120°C under vacuum overnight.

### 1.1.4 Synthesis of RhB@UiO-66 and fluorescein@UiO-66

The guest@UiO-66 materials were synthesised similarly to the UiO-66 crystals but involving one additional step. On one side, 0.4 mmol (93 mg) of $ZrCl_4$ were dissolved in 10 mL of DMF. On the other side, 0.4 mmol (67 mg) of terephthalic acid, 3 mg of the guest dye (RhB or fluorescein) and 2.8 mL of glacial acetic acid were added to 10 mL of DMF. Both solutions were mixed into a Pyrex bottle, and the mixture was heated in an oven at 120 °C for 24 h. The yellow (fluorescein) and pink (RhB) powder samples were thoroughly washed with DMF and MeOH until two consecutive supernatants were transparent and non-emissive under UV irradiation (this typically involves between 5-7 washing steps). After that, the powder samples were collected by centrifugation (8000 rpm), and dried at 90 °C under vacuum overnight.

## 1.2 MATERIALS CHARACTERISATION

### 1.2.1 PXRD and FTIR

Powder X-ray diffraction (PXRD) and FTIR confirmed the successful synthesis of ZIF-8 and UiO-66. PXRD patterns were measured using the Rigaku MiniFlex diffractometer equipped with a Cu Kα source (step size of 0.02° and 0.01°/min) and validated against the simulated XRD pattern. Attenuated total reflection (ATR)-FTIR measurements on the (bulk) polycrystalline powder material were performed using the Nicolet iS10 FTIR spectrometer. Scanning electron microscope (SEM) images of the samples were obtained with a TESCAN LYRA3 electron microscope.

### 1.2.2 NanoFTIR Measurements

The near-field optical measurements were performed using the neaSNOM instrument (neaspec GmbH) based on a tapping-mode AFM where the platinum-coated tip (NanoAndMore GmbH, cantilever resonance frequency 250 kHz, nominal tip radius ~20 nm) was illuminated by a broadband infrared (IR) laser. The coherent mid-infrared (MIR) light (700-2400 $cm^{-1}$) was generated through the nonlinear difference-frequency combination of two beams from fibre lasers (TOPTICA Photonics Inc.) in a GaSe crystal. By demodulating the optical signal at higher harmonics of the tip resonance frequency and employing a pseudo-heterodyne interferometric detection module, amplitude and phase of the scattered wave from the tip are measured, hence, yielding the complex optical response of the material without background signals.[2] We probed individual MOF crystals by averaging over three measurements with 20 individual point spectra each. At least 20 different crystals or regions were probed for each sample. Each spectrum was acquired from an average of 20 Fourier-processed interferograms with 10 $cm^{-1}$ spectral resolution, 2048 points per interferogram, and an 18-ms integration time. The sample spectrum was normalised to a reference spectrum measured on a gold surface to reconstruct the final nanoFTIR amplitude and phase. The continuous broadband MIR spectra were attained by combining two illumination sources, then the obtained spectra were joined at 1500 cm$^{-1}$. All measurements were carried out under ambient conditions (~40% RH).

### 1.2.3 s-SNOM Imaging

The s-SNOM contrast images were measured at neaspec GmbH (Haar, Germany). As the source of monochromatic irradiation, a tuneable quantum cascade laser (Daylight solutions) was employed with output powers tuned to approximately 2 mW. For each scan, the pixel integration time was set as 16 ms. The tip was operating at a frequency of 253 kHz. The recorded signal was demodulated at the third harmonic through a pseudo-heterodyne detection mode. It is worth mentioning that the measurements were repeated at various positions for a systematic characterisation of the sample, and only exemplary scans are presented in Figures 3 & 5 of the main manuscript.

### 1.2.4 Fluorescence Spectroscopy

Steady-state fluorescence spectra were recorded employing the FS-5 spectrofluorometer (Edinburgh Instruments) equipped with different modules for specific measurements. For RhB@ZIF-8, the powder samples were dispersed in acetone, and the SC-05 module, a standard cuvette holder for liquid samples, was used. Excitation and emission spectra of UiO-66 and Guest@UiO-66 powder samples were obtained using the SC-10 module for solid materials. The detection and excitation wavelengths for measuring the steady-state excitation and emission spectra, respectively, are indicated in the corresponding figures. Each measurement is acquired from two scans with a dwell time of 0.2 s and a step size of 1 nm.

### 1.2.5 Fluorescence Lifetime Imaging Microscopy (FLIM)

The encapsulation of fluorescein into UiO-66 was further corroborated by using a confocal fluorescence microscope. Fluorescence lifetime images (FLIM) of fluorescein@UiO-66 were recorded using an inverted-type scanning confocal fluorescence microscope (MicroTime-200, Picoquant, Berlin) with a 60× NA1.2 Olympus water immersion objective, and a 2D piezo scanner (Physik Instrumente). A 470-nm pulsed diode laser (pulse width ~40 ps) was employed as the excitation source. A dichroic mirror (AHF, Z375RDC), a 500-nm long-pass filter (AHF, HQ500lp), a 75-µm pinhole, and an avalanche photodiode detector (MPD, PDM series) were used to collect the emission. The emission spectrum was recorded using a spectrograph (Andor SR 303i-B) equipped with a 1600 × 200 pixel EMCCD detector (Andor Newton DU-970N-BV) coupled to the Micro-Time-200 system.

Note that only the fluorescein@UiO-66 crystals can be probed under this FLIM system using the available excitation source of 470 nm. Unfortunately, RhB cannot be studied by the current setup because this will require an excitation source of at least 520 nm.

### 1.2.6 Density Functional Theory (DFT) Calculations

The theoretical vibrational spectrum of ZIF-8 was computed using DFT at the PBE level of theory[3] with a damped empirical dispersion term (PBE-D).[4] The calculations were carried out with the periodic *ab initio* CRYSTAL14 code.[5] After a geometry optimisation, the Berry Phase approach[6] was employed to compute the IR intensities. A continuous spectrum was obtained by fitting the calculated IR intensities with Lorentzian peak shapes with a FWHM of 10 cm$^{-1}$.

DFT calculations to determine the theoretical vibrational spectrum of RhB were performed using GAUSSIAN 09 software.[7] Both the geometry optimisation and final frequency calculations were carried out by employing the BLYP/6-311+G(2d,p) basis set.

## 2  FURTHER DETAILS: MEASURING THE LOCAL INFRARED ABSORPTION SPECTRA

The processed s-SNOM data yield the optical properties of the sample at a nanoscale (henceforth we refer to such properties as 'local'). Derived from the standard theoretical description of s-SNOM, the scattering coefficient relates the incident field with the scattered field from the tip. Since the backscattered light from the tip is modified by the near-field interaction with the sample, the complex-valued scattering coefficient contains information about the sample permittivity. However, the signal is dominated by the background scattering. To extract the near-field interaction, the detector signal is modulated at higher harmonics of the tapping amplitude of the AFM tip. Demodulation of the interferometric detector signal is achieved by a complex Fourier transform with respect to time. Here, it is worth mentioning that the model for the scattering coefficient includes the tip height above the sample surface, which depends on time due to the oscillation of the tip.

The Fourier transformation yields the sample dielectric function, from which the local permittivity can be extracted, for a detailed mathematical description we refer to Govyadinov et al.[8] Note that the measured data includes contributions from the setup, hence normalisation to a reference signal is required. This provides the complex-valued near-field contrast with its corresponding amplitude and phase. However, the imaginary part of the near-field contrast obtained through nanoFTIR depicts a better match with the far-field FTIR absorption spectra,[8] where the absorption coefficient is attained *via* the Kramer's-Kronig relations.[9] Therefore, the nanoFTIR absorption is defined as the imaginary part of the near-field contrast.

Hence, only the imaginary part is shown in this work. For completeness, Figure S1 summarises the raw data of the amplitude, phase, as well as the real and imaginary parts measured on an individual ZIF-8 crystal, and normalised to the gold substrate. The continuous broadband MIR spectrum is attained by combining two illumination sources. First, Laser 1 (700 – 1400 $cm^1$) was employed to measure the reference spectra on the clean gold substrate, acquired from an average of 20 interferograms. Each interferogram was measured with a 10 $cm^{-1}$ spectral resolution, 2048 points per interferogram and with an 18-ms integration time.

Secondly, the ZIF-8 crystal, for instance, was probed to attain the nanoFTIR spectra, which were divided by the reference spectra. To reduce the noise, three measurements on the same location of the crystal were performed and averaged. The same procedure was repeated after switching the IR illumination source to Laser 2 (1250 – 2100 cm$^{-1}$). As shown in Figure S1, the spectra obtained from the same crystal matched well in the overlapping region and can thus be combined to yield the broadband spectra from 700 to 2100 cm$^{-1}$.

Figure S2 depicts exemplary positions, where the local nanoFTIR spectra were measured. To examine the IR spectra from the MOF materials, at least 20 individual crystals were measured and compared to obtain information about the local chemical variability of the individual crystals as well as to determine the properties of the sample, when averaged over the measurements on several single crystals.

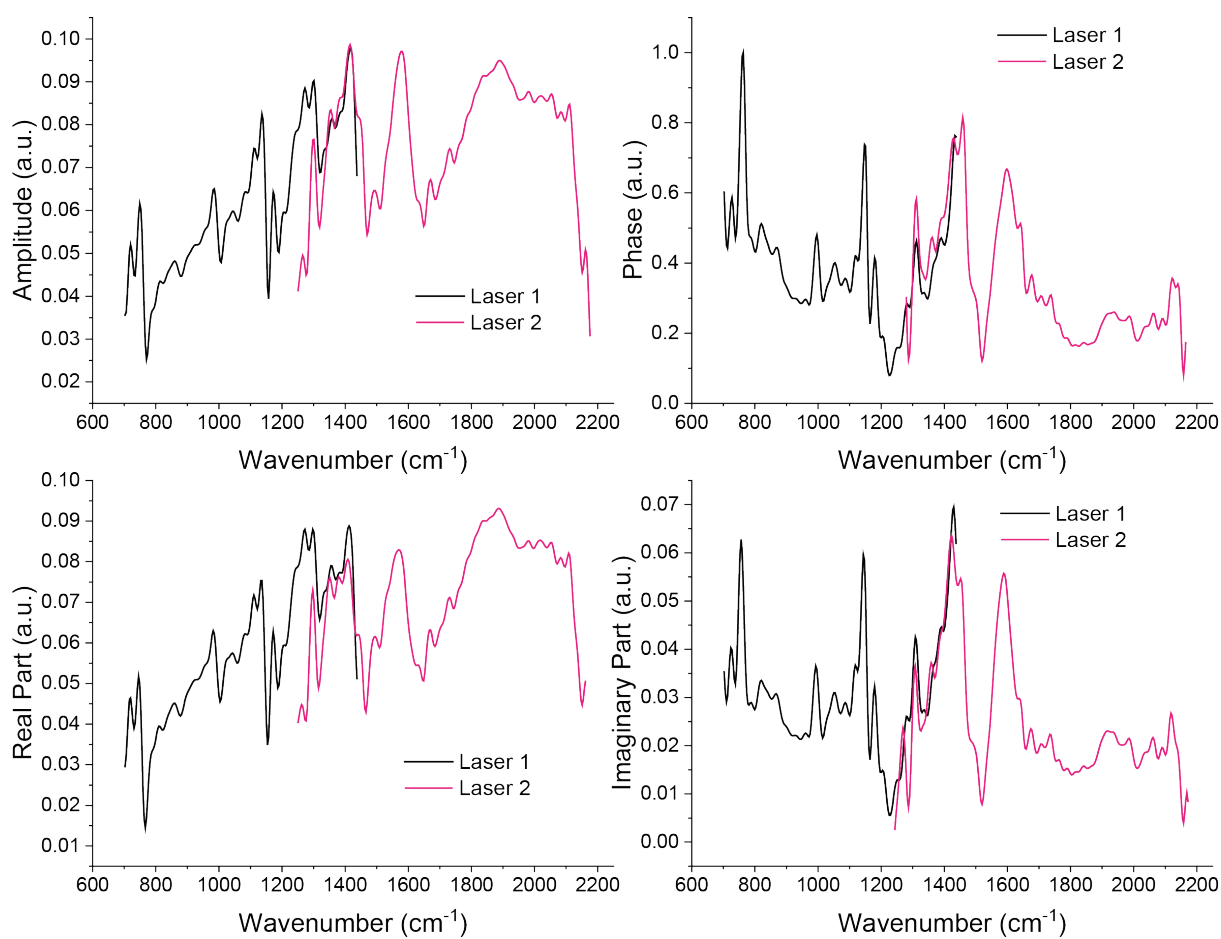

**Figure S1:** Amplitude, phase, real and imaginary parts of the near-field spectra obtained through nanoFTIR measurements on a single crystal of ZIF-8, normalised to a gold substrate.

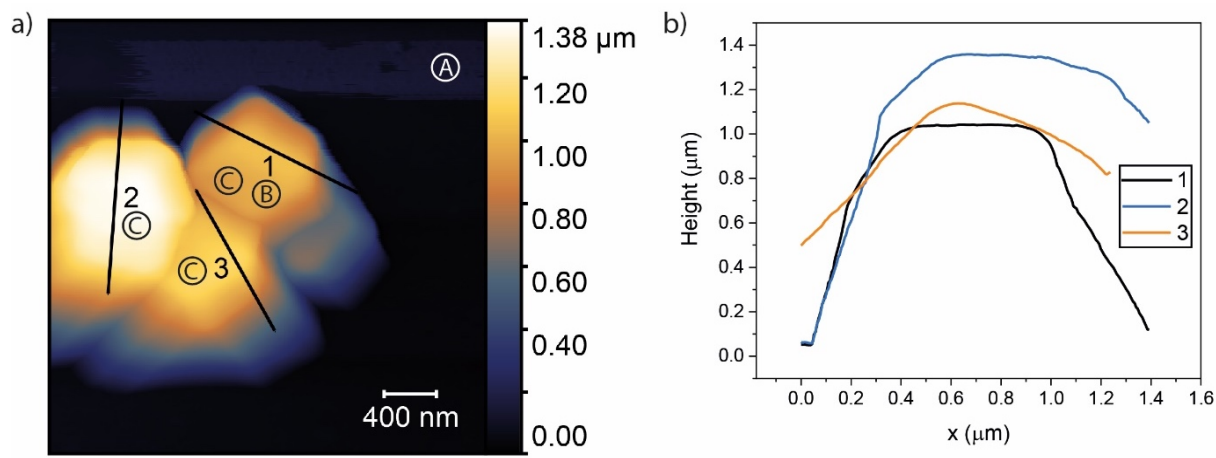

**Figure S2:** a) The exemplary positions of the local nanoFTIR measurements performed on micron-sized crystals. Region A: reference spectrum measured on silicon or gold substrate. Region B: Three measurements with 20 individual point spectra each were performed to obtain the local properties of an individual crystal. Regions C: Several crystals were measured (at various positions on the sample, not shown here) and further compared to probe the sample-specific properties. b) Height profiles of the crystals along the designated lines (1-3) in the AFM image.

## 3 FURTHER DETAILS: MATERIALS CHARACTERISATION

### 3.1 X-RAY DIFFRACTION (XRD)

XRD was employed to confirm the crystalline structure of the powder samples. Through comparison of the experimental measurements with simulated reference patterns, the successful synthesis of the MOF crystals was validated as shown in Figures S3 & S4. The crystal structure of the RhB@ZIF-8 system is in agreement with the simulated ZIF-8 pattern. Figure S4 reveals the intact crystalline structure of the guest@UiO-66 systems (guest = rhodamine B or fluorescein), matching the diffraction pattern of the pristine UiO-66 crystal. The reference XRD patterns were analysed using the CrystalDiffract software, using CIF files from the Cambridge Structural Database (CSD) database as input.

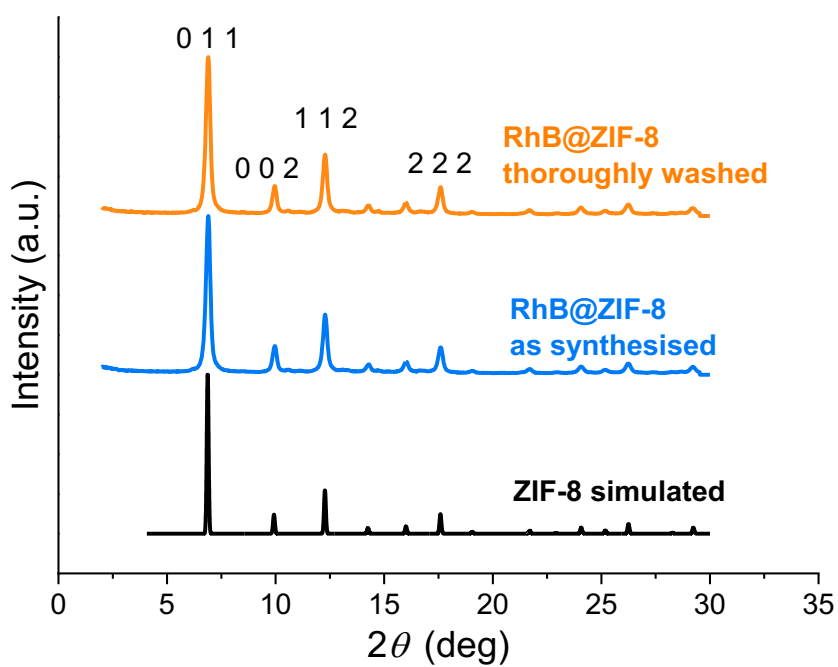

**Figure S3:** XRD patterns of RhB@ZIF-8 powder samples compared with the simulated pattern of ZIF-8.

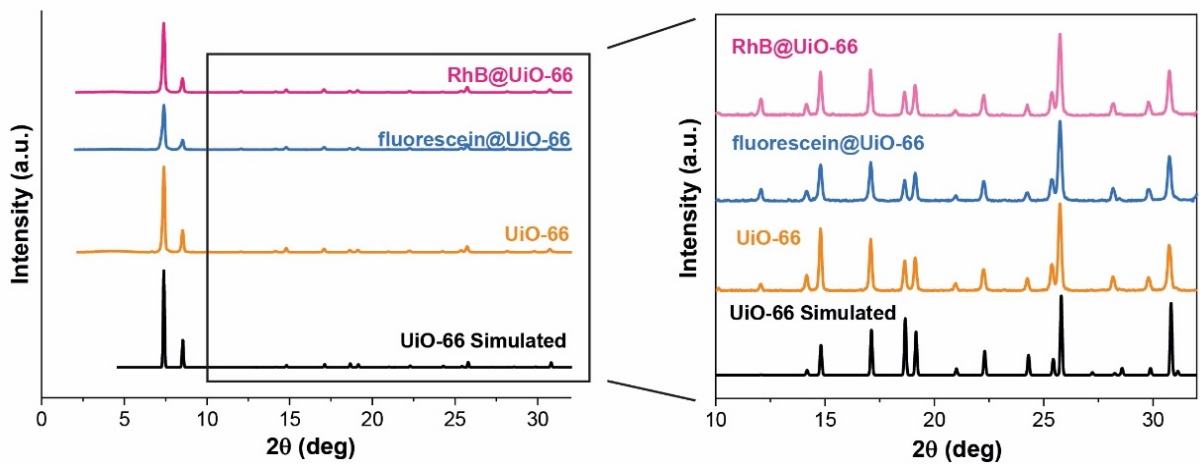

**Figure S4:** XRD patterns of the guest@UiO-66 composite systems.

## 3.2 ATOMIC FORCE MICROSCOPY (AFM)

AFM yields a topographical image of a sample surface at nanometre resolution. In addition, the roughness of a surface or the thickness of a crystal can be quantified. AFM imaging was performed with a neaSNOM instrument (neaspec GmbH) operating in tapping mode. Height topography images were collected using the Scout350 probe (NuNano), which has a nominal tip radius of 5 nm, a spring constant of 42 N/m and resonant frequency of 350 kHz. Figures S5 – S7 exemplarily demonstrate height profiles of the MOF crystals under investigation. Further, the AFM measurements were combined with nanoFTIR spectroscopic data to yield the combined topographical and chemical information at the nanoscale.

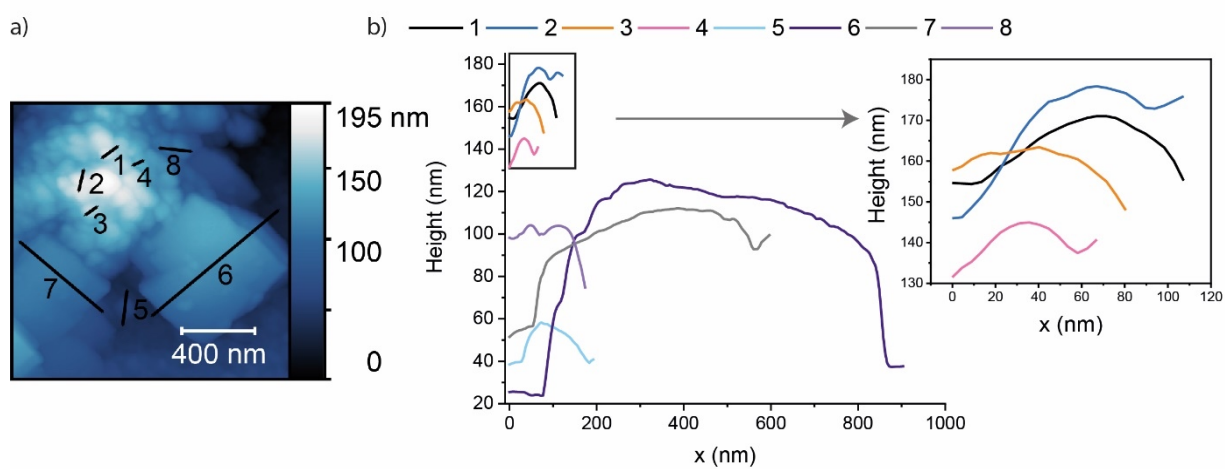

**Figure S5:** (a) AFM image of the ZIF-8 nanocrystals and RhB blocks. (b) Surface height profiles corresponding to the designated lines in (a).

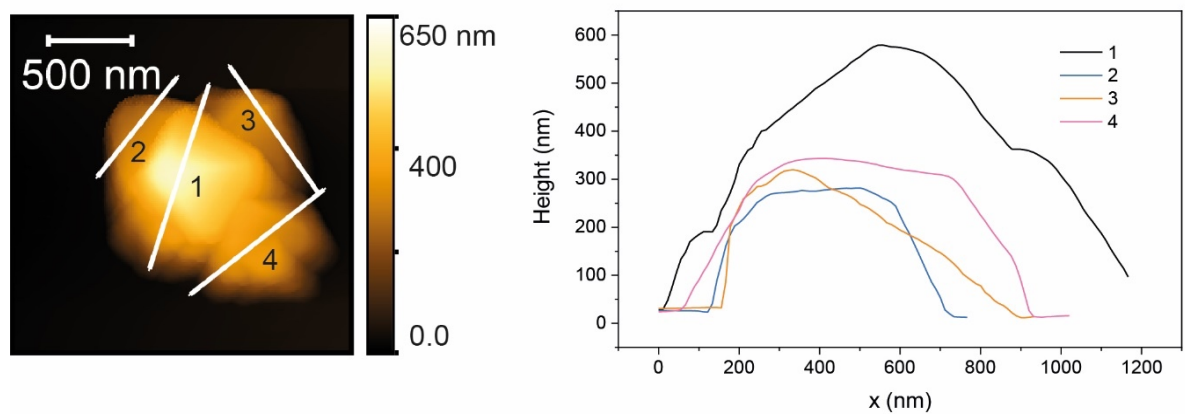

**Figure S6**: Height profiles of the fluorescein@UiO-66 crystals.

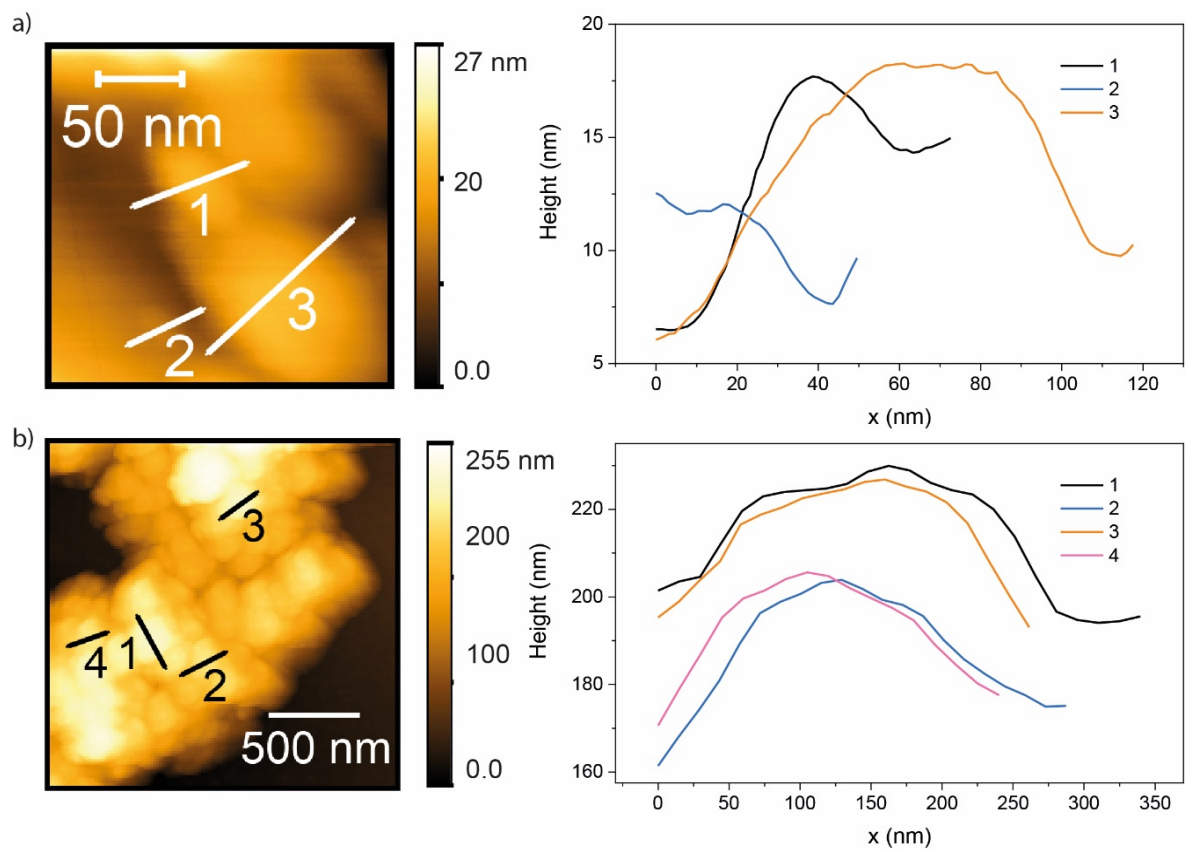

**Figure S7:** Height profiles of RhB@ZIF-8 nanocrystals: (a) as-synthesised and (b) after thorough washing.

### 3.3 SEM Imaging

Backscattered electron and secondary electron SEM images were obtained at 10 keV under high vacuum. SEM characterisation was not only performed to image the MOF-type crystals, but to further confirm the absence of any residual guest material on the sample surface. The representative images of the samples at various locations are shown in Figures S8 and S9.

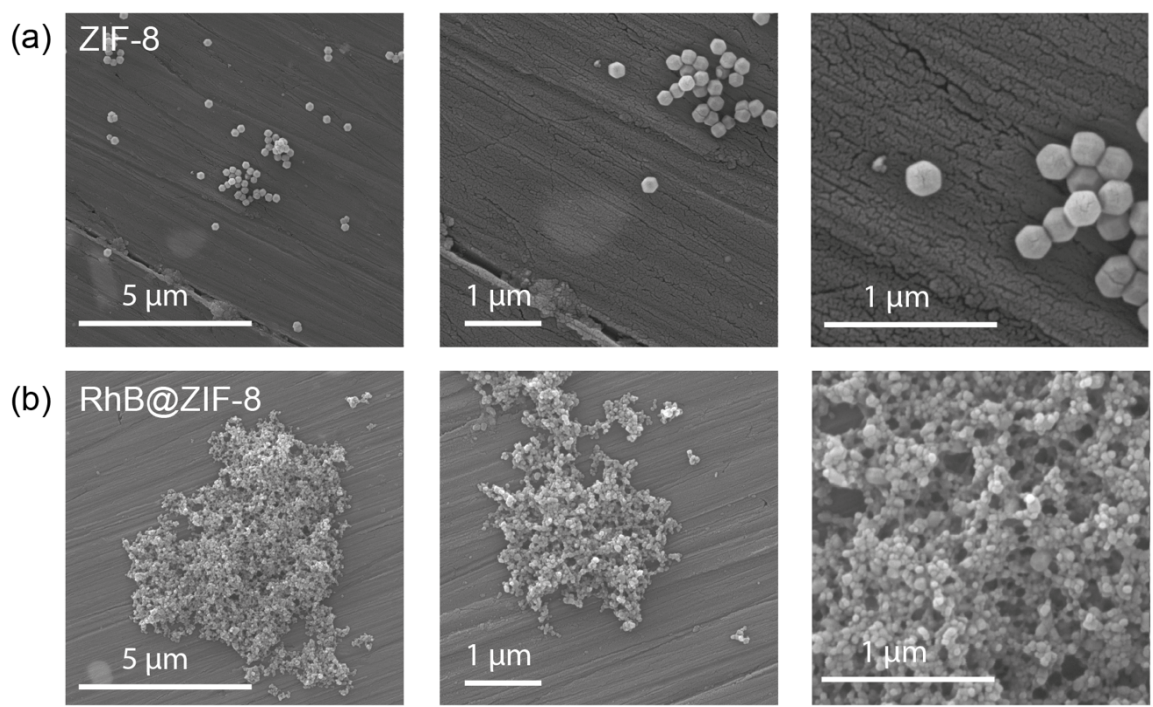

**Figure S8:** SEM micrographs of pristine (a) ZIF-8 and (b) RhB@ZIF-8 crystals.

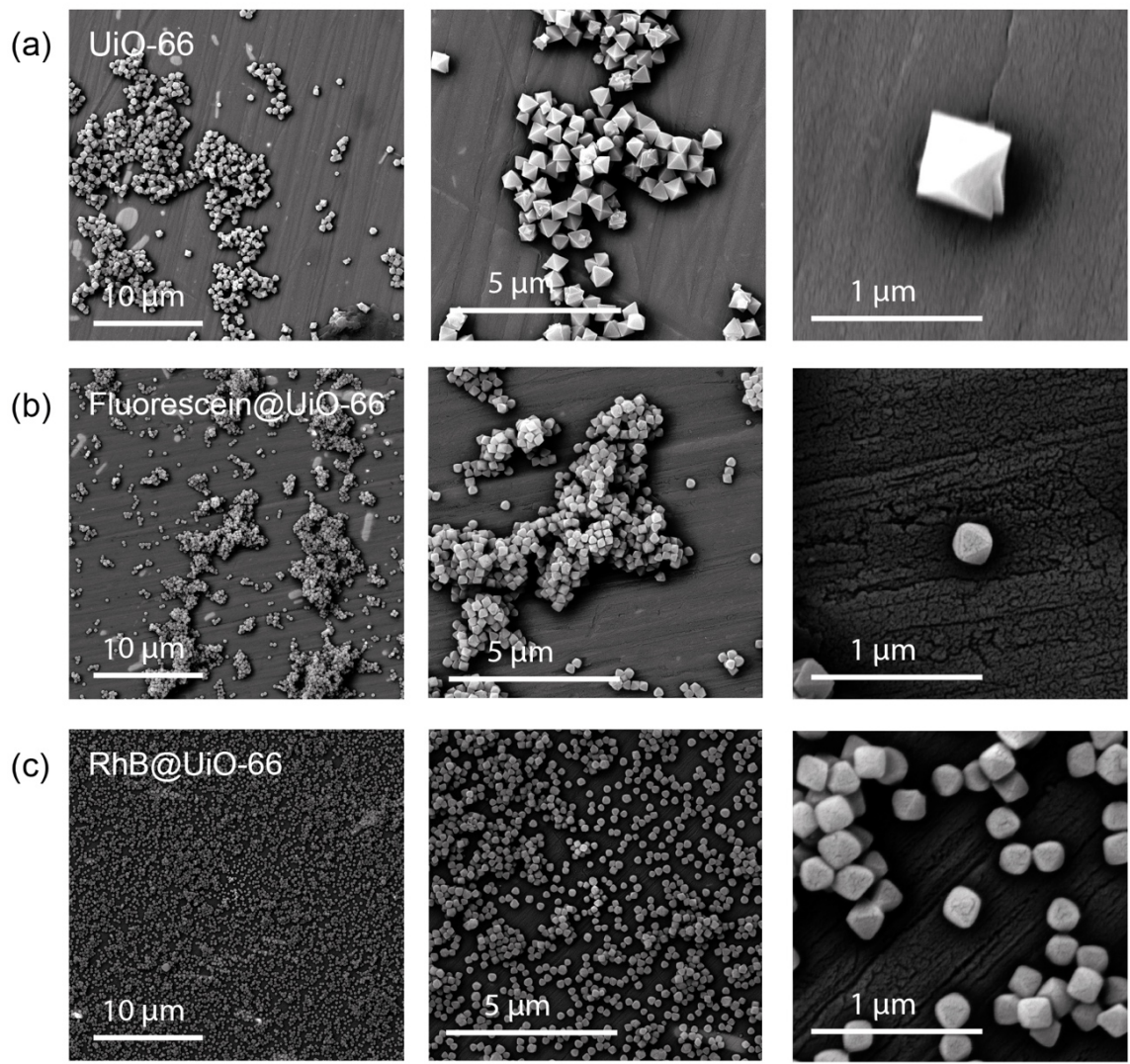

**Figure S9:** Backscattered electron SEM micrographs of the single crystals of (a) UiO-66, (b) fluorescein@UiO-66, and (c) RhB@UiO-66.

### 3.4 FLUORESCENCE LIFETIME IMAGING MICROSCOPY (FLIM)

Figure S10 shows that the lifetime of the fluorescein@UiO-66 single crystals is a monoexponential decay of 3.77 ns, which correlates well the lifetime observed for fluorescein in different solvents. Fluorescence lifetime images of three different crystals of UiO-66 are shown in Figure S11.

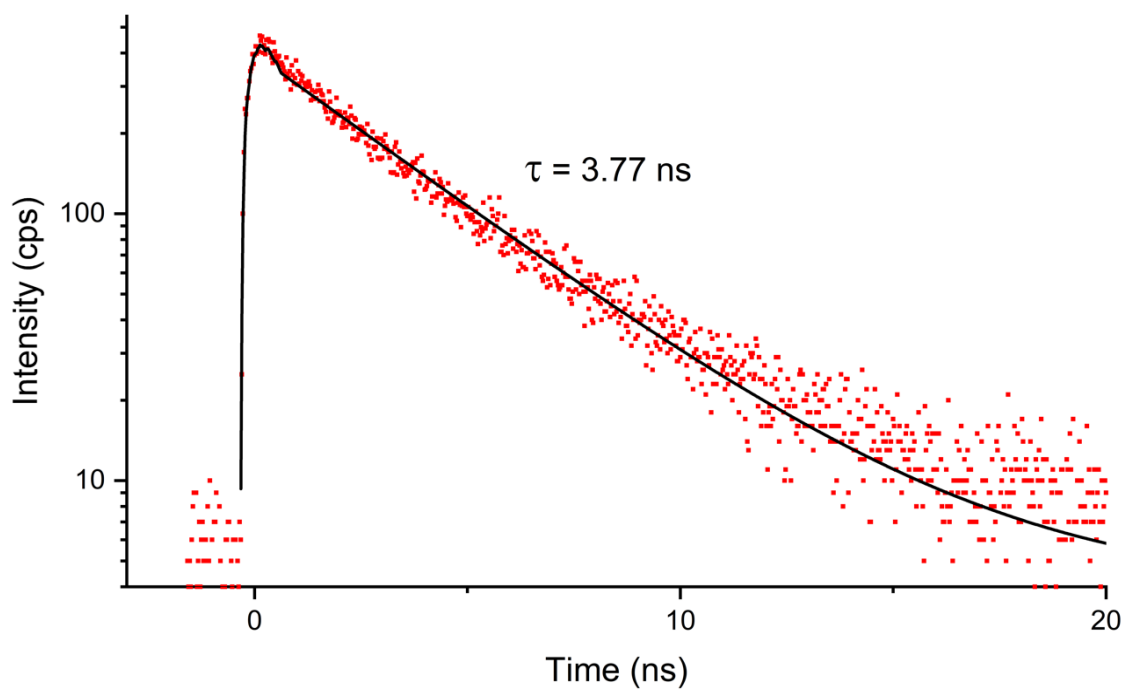

**Figure S10:** Average luminescence decay (lifetime) curve of several fluorescein@UiO-66 single crystals with mean lifetime of 3.77 ns (standard deviation: 0.15).

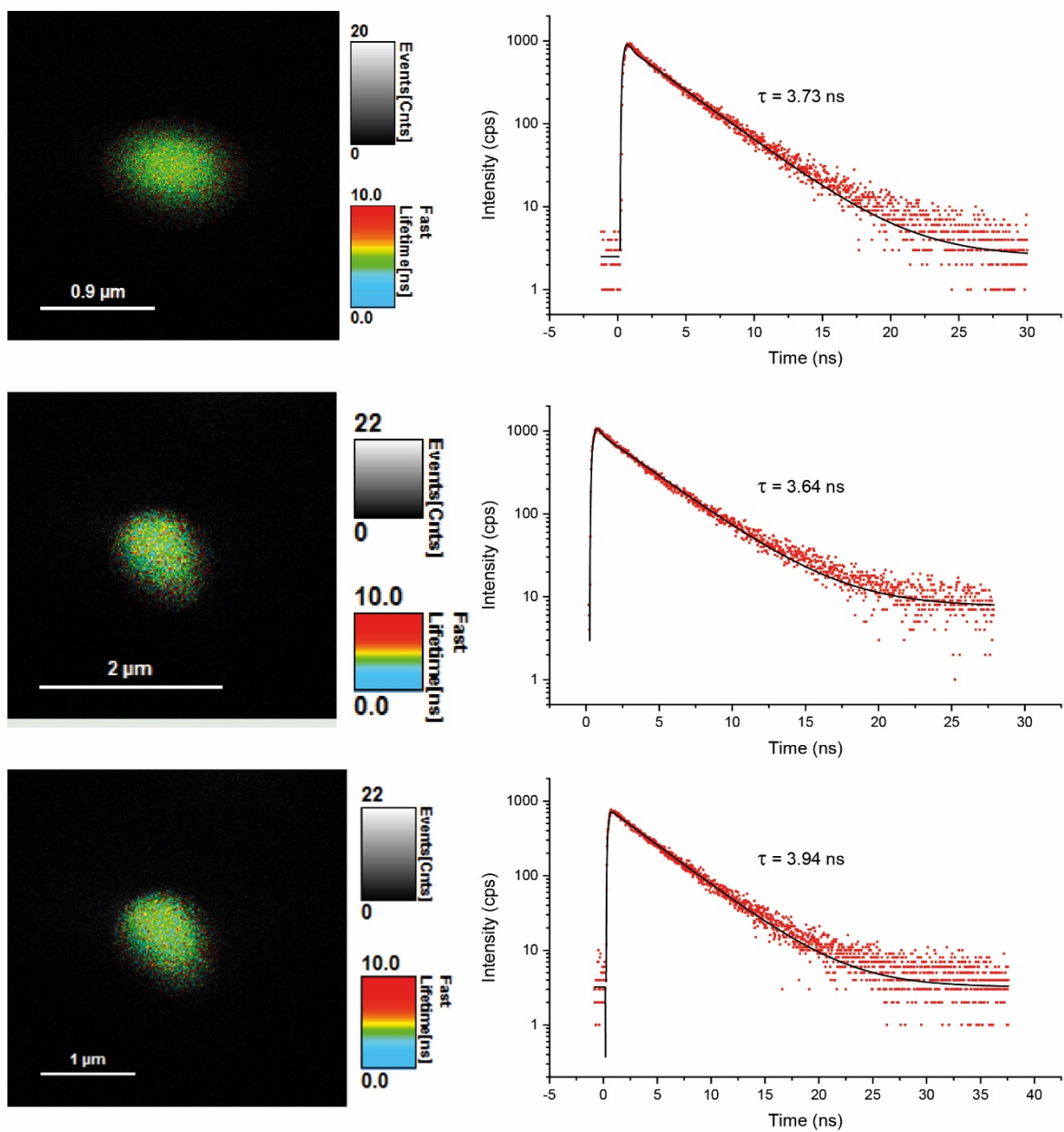

**Figure S11:** Fluorescence lifetime images and decay curves of fluorescein@UiO-66 single crystals (false colour scale).

## 3.5 WASHING PROCESS

The effectiveness of thorough washing has been confirmed by probing the supernatant of the last washing step. If any trace of the luminescent dye is still detectable, then the washing process has not been completed. However, due to the absence of any guest material found in the supernatant, we can conclude that the previous washing was thorough enough to remove any guest material. This is demonstrated by the absorption spectra of the supernatant – DMF in the case of Guest@UiO-66 and acetone for RhB@ZIF-8 – which show no signal of the guest (Fig. S12 a,c). In addition, emission spectra, with their higher sensitivity to the dyes, were measured and compared with the obtained Guest@MOF material (Fig. S12 b,d). Again, we found no sign of any guest material in the supernatant.

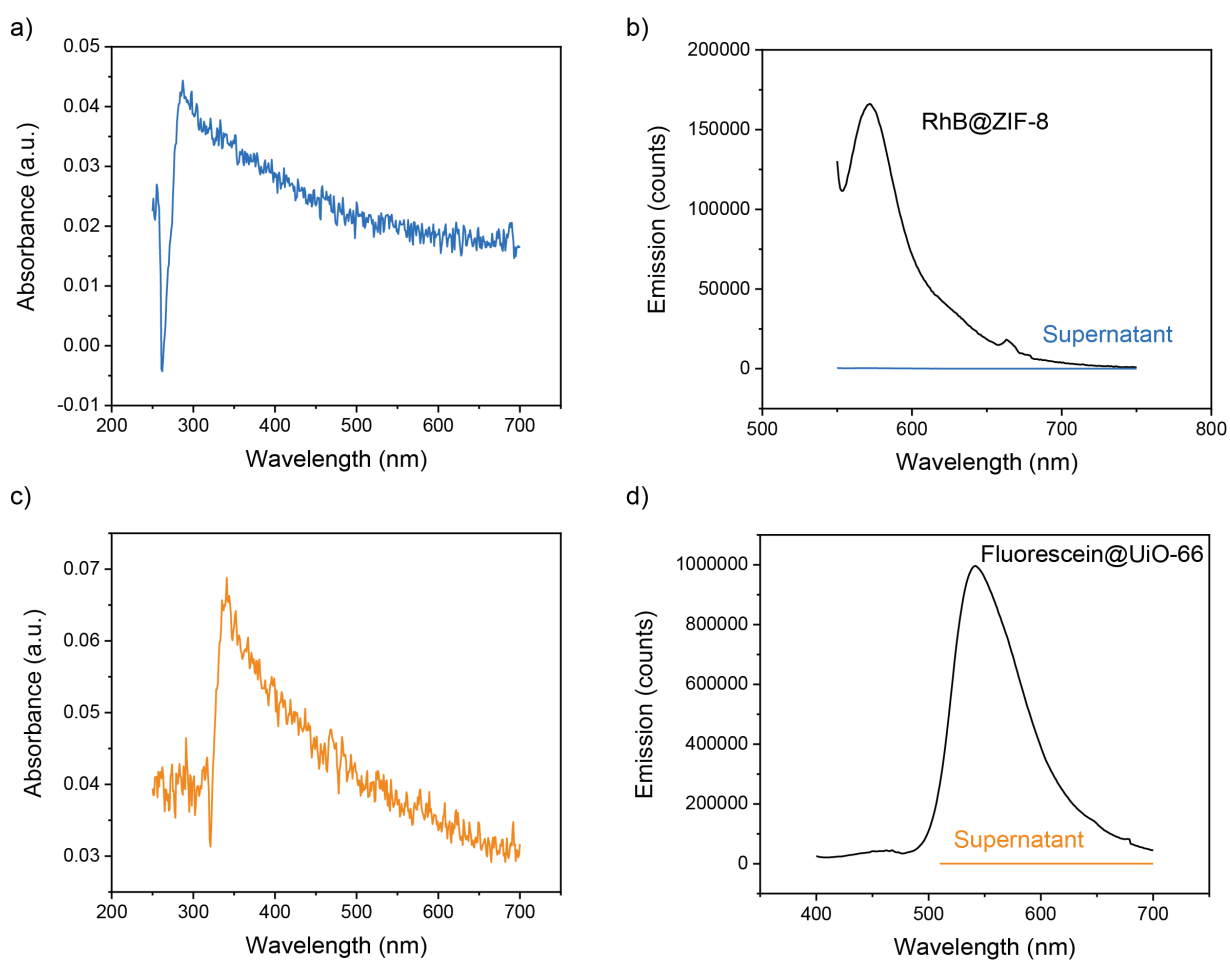

**Figure S12:** Absence of the dyes in the supernatant of the last washing step confirming thorough washing. a) Absorbance spectrum of acetone supernatant after removal of RhB@ZIF-8 showing no signal of RhB. b) Emission spectrum of acetone supernatant compared with the RhB@ZIF-8 (excitation wavelength 540 nm). c) Absorbance spectrum of DMF supernatant after removal of fluorescein@UiO-66 showing no signal of the fluorescein

dye. d) Emission spectrum of DMF supernatant (excitation wavelength 500 nm) compared with the fluorescein@UiO-66 (excitation wavelength 380 nm).

## 4 REFERENCES

1. Su, Z.; Miao, Y.-R.; Zhang, G.; Miller, J. T.; Suslick, K. S. Bond breakage under pressure in a metal organic framework. *Chem. Sci.* **2017,** 8 (12), 8004-8011.
2. Keilmann, F.; Hillenbrand, R. Near-field microscopy by elastic light scattering from a tip. *Phil. Trans. R. Soc. A* **2004,** 362 (1817), 787-805.
3. Perdew, J. P.; Burke, K.; Ernzerhof, M. Generalized Gradient Approximation Made Simple. *Phys. Rev. Lett.* **1996,** 77 (18), 3865-3868.
4. Grimme, S. Semiempirical GGA-type density functional constructed with a long-range dispersion correction. *J. Comput. Chem.* **2006,** 27 (15), 1787-1799.
5. Dovesi, R.; Orlando, R.; Erba, A.; Zicovich-Wilson, C. M.; Civalleri, B.; Casassa, S.; Maschio, L.; Ferrabone, M.; De La Pierre, M.; D' Arco, P.; Noël, Y.; Causà, M.; Rérat, M.; Kirtman, B. CRYSTAL14 : A program for the ab initio investigation of crystalline solids. *Int. J. Quantum Chem.* **2014,** 114 (19), 1287-1317.
6. Noel, Y.; Zicovich-Wilson, C. M.; Civalleri, B.; D'Arco, P.; Dovesi, R. Polarization properties of ZnO and BeO: An ab initio study through the Berry phase and Wannier functions approaches. *Phys. Rev. B* **2001,** 65 (1), 014111.
7. Frisch, M. J., et al. *Gaussian 09, Revision A.02,* , Gaussian, Inc.: Wallingford CT, 2009.
8. Govyadinov, A. A.; Amenabar, I.; Huth, F.; Carney, P. S.; Hillenbrand, R. Quantitative Measurement of Local Infrared Absorption and Dielectric Function with Tip-Enhanced Near-Field Microscopy. *J. Phys. Chem. Lett.* **2013,** 4 (9), 1526-1531.
9. Chalmers, J. M.; Griffith, P. R., *Handbook of Vibrational Spectroscopy*. John Wiley & Sons, Ltd: 2006; Vol. 1.